\documentstyle[aps,twocolumn,epsf,psfig]{revtex}

\begin{document}

\twocolumn[\hsize\textwidth\columnwidth\hsize\csname @twocolumnfalse\endcsname



\title{Coupling    between   phonons   and   intrinsic   Josephson 
oscillations in cuprate superconductors}

\author{Ch.~Helm,  Ch.~Preis, F.~Forsthofer  and J.~Keller,  \\ 
Institut  f\"ur Theoretische  Physik,  Universit\"at  Regensburg, 
D-93040 Regensburg,  Germany \\ K.  Schlenga, R.  Kleiner, and P. 
M\"uller,   \\   Physikalisches   Institut   III,   Universit\"at 
Erlangen-N\"urnberg, D-91058 Erlangen, Germany} 

\date{\today, submitted to Physical Review Letters}

\maketitle

\begin{abstract}  
The  recently  reported  subgap  structures      in  the 
current-voltage characteristic of 
intrinsic Josephson  junctions  in the high-$T_c$ superconductors 
Tl$_2$Ba$_2$Ca$_2$Cu$_3$O$_{10+\delta}$                       and 
Bi$_2$Sr$_2$CaCu$_2$O$_{8+\delta}$  are explained by the coupling 
between  c-axis phonons and Josephson  oscillations.   A model is 
developed  where c-axis  lattice  vibrations
between adjacent superconducting multilayers are excited  by the 
Josephson oscillations in a resistive junction. The voltages 
of the lowest structures  correspond  well to the frequencies  of 
longitudinal  c-axis phonons  with large oscillator  strength 
in the two materials, providing a new measurement technique for this
quantity.
\end{abstract}   

]

The  transport  properties  of   highly  anisotropic  cuprate 
superconductors  in  c-direction  can  well  be  described  by  a 
stack  of  Josephson  junctions  formed  by  non-superconducting  material 
between adjacent superconducting copper oxide multi-layers \cite{kleiner}.
Recently the observation 
of  subgap  structures  in  the  current-voltage ($I$-$V$)-characteristic
  of  intrinsic 
Josephson    junctions    in   the    high-$T_c$    superconductors 
Tl$_2$Ba$_2$Ca$_2$Cu$_3$O$_{10+\delta}$        (TBCCO)        and 
Bi$_2$Sr$_2$CaCu$_2$O$_{8+\delta}$ (BSCCO) has been reported 
\cite{yurgens,schlenga,seidel}. 
Each  individual   branch   of  the $I$-$V$-curve shows  a 
structure      which  can  be  traced  back  to  the  
$I$-$V$-characteristic of one single Josephson junction 
in the resistive  state.  
These structures seem to be an intrinsic effect, as they have been observed 
both in step edge junctions
(TBCCO) and mesa-type stacks (BSCCO) of different sizes. 
The characteristic voltages are completely independent of temperature and 

external magnetic fields, which rules out any relation to the superconducting 
gap, vortex flow or the thermal excitation of quasiparticles.

It was shown that the pattern  of one junction  can be 
described phenomenologically 
 by a resistively  shunted  junction  (RSJ) model by 
assuming ad hoc  a   special   structure   for   the   current-voltage 
characteristic of the quasiparticles \cite{schlenga}. 
It was argued that such a 
structure might result from peaks in the quasiparticle density of 
states due to Andreev reflection between normal and superconducting regions.
Several alternative approaches, including the modulation of 
the tunneling distance due to Raman-active phonons, have been mentioned 
in \cite{yurgens}, but up to now all suggestions 
failed to explain the main features of the effect.
In this paper we want to discuss a different mechanism involving phonons  by 
assuming  that the local electric field oscillations  
produced  by the Josephson 
effect in a single  junction excite infrared active c-axis phonons.   In the 
following  we will present a simple model where  we couple the 
non-linear   current-phase   relation   of  one  junction  to  a local 
oscillator. 
The analytical and numerical solution of this model provides a very good 
quantitative explanation of the experimental data.
It is shown that the peaks 
in  the subgap  structure  of  the  dc  current-voltage   characteristic 
correspond  to zeros of the dielectric  function of the barrier 
material,  i.e.  to  longitudinal  optical phonons.

In the minimal version of the 
RSJ model  the total current (density) 
\begin{equation}
i = j_c \sin \gamma + \sigma_0 E + {\dot D}  \; , 
\label{rsjmicro}
\end{equation}
through  one of 
the  junctions  is the sum of the Josephson current,  the (ohmic)
quasiparticle   current  $I_{\rm qp} = \sigma_0 E$
and  the  displacement   current density ${\dot D}$, 
where the (gauge invariant)  phase difference  $\gamma$ is related 
to the   electric  field  $E$ in the barrier  of thickness 
$b$ by 
\begin{equation}
\hbar \dot \gamma = 2eEb \label{josephson} \; .
\end{equation}
Further time dependencies of a microscopic model
 are thereby neglected for simplicity \cite{barone,likharev}.

The  displacement current ${\dot D}$ contains  the  polarization  $P$  of  the 
barrier  medium,  $D=  \epsilon_0  E + P =\epsilon_0 \epsilon E$.
In the case of high frequency Josephson oscillations  in the 
range of phonon frequencies it is important  to keep the full frequency 
dependence of $\epsilon (\omega)$ or to treat the polarization $P$ 
 as an additional dynamical variable. 

Here we assume that the polarization $P = nqz$ is due to a 
c-axis displacement $z$ 
of ions  with  charge  $q$, mass $M$ and density $n$ in the insulating barrier
between  the copper  oxide (multi)-layers.
For  the  motion  of  the  ions  we  assume  a simple  oscillator 
\begin{equation}
\ddot   z  +  \omega^2_0   z  +  r  \dot   z  =  {q \over   M}   E ,
\label{oscillmicro}
\end{equation} 
which is driven by the electric field $E$ in the barrier. 
In this model the contribution  of 
the oscillator to the dielectric function is given by 
\begin{equation}
\epsilon_{\rm ph} ( \omega )  = 
 1 + \frac{n q^2}{\epsilon_0 M} \frac{1}{\omega_0^2 -\omega^2 + i r \omega}
\; .
\end{equation}


It  is  useful  to introduce  normalized  quantities:
   First equ. (\ref{rsjmicro}) and (\ref{oscillmicro}) are  divided  by the 
critical  current  density $j_c$,  then a characteristic  time variable 
$\tau  =  t  \omega_c$  with  $\omega_c  =  (2e/\hbar)   V_c$  is 
introduced,   where   $V_c  =  R  I_c  =  bj_c / \sigma_0$   is  the 
characteristic  voltage determined experimentally  by the voltage 
on the resistive branch at the critical current. 
Then we obtain
\begin{eqnarray}
j &=& \sin \gamma + \dot \gamma + \beta_c \ddot \gamma + \dot p \; , 
        \label{rsjext}\\
\lambda  \dot \gamma &=& \ddot  p + \Omega^2  p + \rho  \, \dot  p \; ,
\label{oscillren}
\end{eqnarray}
with  the normalized  polarization $p= P\omega_c / j_c$, 
  frequency $\Omega = \omega_0/\omega_c$, friction 
$\rho = r/\omega_c$ and 
 the   McCumber para\-meter   $\beta_c$ =  $RC \omega_c$   = 
$\omega_c^2/\omega_J^2$,  where   $\omega_J^2  = 
2ebj_c/(\hbar\epsilon_0)$  is the square of the Josephson  plasma 
frequency  and  $C=  \epsilon_0  F/b$  is the capacitance  of the barrier.
 The coupling constant
$\lambda  = \omega^2_{\rm ion}  /\omega_J^2$ = $S (\omega_0^2 / \omega_J^2)$
can be expressed by the ionic  plasma 
frequency $\omega^2_{\rm ion} = nq^2/(M\epsilon_0)$ or the oscillator 
strength $S$ of the lattice vibration. 
In  the  present   case  the  ratio  between the  phonon   frequency
$\omega_0$  and the Josephson plasma frequency  $\omega_J$  is large;
therefore  even  a small  oscillator  strength  $S = ( \omega_{\rm ion}^2 / 
\omega_0^2 )$ can lead to a sizable coupling.  

The  equations (\ref{rsjext}) and (\ref{oscillren})
 for the phase and the lattice 
displacement  can be solved numerically to yield $\gamma(t)$ 
as a function  of the current  density  $j$  through  the junction. 
Taking  a time  average  of the  phase  the  dc  current-voltage 
characteristic is obtained. 

From the numerical results it turns out that both the phase $\gamma (t)$ 
and the polarization $p(t)$ oscillate primarily with one frequency, which is 
in agreement with general expectations for the RSJ-model with 
 large $\beta_c$ \cite{likharev}.
Therefore it is satisfied to neglect 
 the higher harmonics in the ansatz
\begin{eqnarray}
\gamma & = & \gamma_0 + vt + \gamma_1 \sin\omega t, \\
p  &=& p_0 + p_1 \cos (\omega t + \varphi) . 
\end{eqnarray}
Here $v$ is the time averaged phase velocity, which corresponds to the dc 
voltage $v = < V> / V_c $ in the resistive state.

The different Fourier components  of the equations  of motion are 
obtained by the Bessel function expansion 
\begin{equation}
\sin    \gamma(t)    =   \sum_{n=-\infty}^\infty    J_n(\gamma_1) 
\sin(\gamma_0 + vt + n\omega t)  \; .
\end{equation}
The  Josephson  current  contributes  to the dc current  only  if 
$v  + n \omega =0$.

For the  fundamental harmonic we obtain $\omega = v$.
Using this ansatz in the differential  equations (\ref{rsjext}) and 
(\ref{oscillren}) we obtain a set 
of  equations  for  the  amplitudes   of  the  different  Fourier 
components. 
As equ. (\ref{oscillren}) is linear in the polarization $p$, 
we can eliminate the polarization $p$ and 
get an equation of motion for the phase oscillation alone:
\begin{eqnarray}
j &=& v(1 + {1\over 2} \sigma (v) \gamma_1^2), \label{gleich1}  \\
\left( J_0(\gamma_1)   -  J_2(\gamma_1)  \right) \cos\gamma_0 &=&  v^2 
\beta_{\rm eff}(v)\gamma_1, \label{gleich2} \\
J_1(\gamma_1) \sin\gamma_0 &=& - \frac{1}{2} v \sigma (v) \gamma^2_1  
\label{gleich3}
\end{eqnarray} 
with 
\begin{eqnarray}
\beta_{\rm eff}(v) &=& \beta_c - {\lambda (v^2-\Omega^2)\over
(v^2-\Omega^2)^2 + v^2\rho^2},\\
\sigma(v) &=& 1 +  {\lambda v^2 \rho\over
(v^2-\Omega^2)^2 + v^2\rho^2} \; .
\end{eqnarray}

The  functions  $\beta_{\rm eff}(v)$  and $\sigma (v)$  are related  to the 
real  and  imaginary  part  of the phonon dielectric  function 
$ \epsilon_{\rm ph} ( \omega ) = \epsilon_1 + i \epsilon_2$  via 
\begin{eqnarray}
\beta_{\rm eff}(v)  &=& \beta_c  \,  \epsilon_1,\\
\sigma (v) &=& 1 + v  \beta_c \,  \epsilon_2 .
\end{eqnarray}
Thereby the poles (zeros)  of $\epsilon (\omega) $ are given by the 
eigenfrequencies $\omega_{\rm TO}$ 
($\omega_{\rm LO}$) of transversal (longitudinal) optical phonons for wavevectors 
${\vec k} =0$.
In this way the formalism can be easily extended to an arbitrary number of 
phonon branches and more complicated lattice dynamical models. 

As the numerical solution of equ. (\ref{rsjext}) and (\ref{oscillren}) 
shows that $\gamma_1 < 0.1$, the equations (\ref{gleich1})-(\ref{gleich3})
can be linearized in $\gamma_1$ and an analytical formula  for the 
$I$-$V$-characteristic can be obtained:
\begin{eqnarray}
j(v) = v + \Delta j (v) &=& 
       v + \frac{1}{2v} \frac{\sigma}{\sigma^2 + {(v \beta_{\rm eff} )}^2}
             \\
      &=& v - \frac{1}{2 v^2} {\rm Im} \left( 
        \frac{1}{\beta_{\rm eff} + i {\sigma \over v} } \right)  \; \; . 
          \label{iv2}
\end{eqnarray}
Higher orders in $\gamma_1$ have been calculated analytically, but have only
a negligible influence on the $I$-$V$ characteristic. 

With the help of these relations some special points of 
the $I$-$V$-characteristic $j(v)$ near the  subgap
structures can be identified:
 
For small voltages $ v \ll \Omega_{\rm TO}= \Omega = {\omega_{\rm TO} 
/ \omega_{\rm c} } $ one has 
$\beta_{\rm eff} (v) \approx \beta_c$ and $ \sigma(v) \approx 1$ and 
the  model reduces to the conventional RSJ-model. For this it is well known 
\cite{barone} that  there is a voltage jump in the $I$-$V$-characteristic 
at $v_{\rm min} \approx 4 \omega_J / ( \pi \omega_c ) $.

At the resonance $v = \Omega_{\rm TO}$ of the phononic oscillator
both the effective quasiparticle conductivity $\sigma (v)$ and the effective 
McCumber parameter $\beta_{\rm eff}$ are strongly enhanced and the 
$I$-$V$ characteristic in equ.  (\ref{iv2})
reduces to the purely ohmic term.  
This corresponds to a pure quasiparticle tunneling 
current across the junction, while the supercurrent and the displacement 
current are compensating each other.

In contrast to this,  equ. (\ref{iv2}) indicates 
 a resonance in $j(v)$
near the zeros of $\beta_{\rm eff}$, i.e. the eigenfrequencies 
$\Omega_{\rm LO}$ of {\em longitudinal} optical phonons. 

The difference $\Delta v := \Omega_{\rm LO} - v_{\rm max}$ 
between $\Omega_{\rm LO}$ and the actual maximum $v_{\rm max}$ of $j(v)$ 
can be estimated as $\Delta v < 2 \%$, 
which is  almost independent of the choice of parameters. 
Physically this point is connected with an oscillation of the 
external electric field $E$ and the polarization $p$  with vanishing 
displacement current density ${\dot D}$.
With the help of equ. (\ref{iv2}) also an analytical formula for the 
intensity 
\begin{equation}
\Delta j_{\rm max} := \Delta j ( \Omega_{\rm LO} ) 
 = \frac{1}{2 \Omega_{\rm LO} \sigma (\Omega_{\rm LO})}
\end{equation}
of the subgap structures can be derived. 
Thereby it turns out that a small damping parameter $\rho$,
corresponding to  a weak coupling of inter-layer ions  in  neighbouring 
contacts or equivalently the small dispersion of the phonons
in c-direction, is crucial for the existence 
of a hysteretic region; no maximum of $j(v)$
can be found for $\rho \ge \rho_{\rm crit} 
= \Omega_{\rm LO} - \Omega_{\rm TO}$. 
Also note that the intensity $\Delta i_{\rm max} = j_{\rm c} \cdot \Delta 
j_{\rm max}$
 in ordinary units is 
proportional to the critical current $j_c$, as it has been reported in 
experiments \cite{schlenga}.

\begin{figure}
\leavevmode
\begin{center}
\epsfxsize=0.45\textwidth
\epsfbox{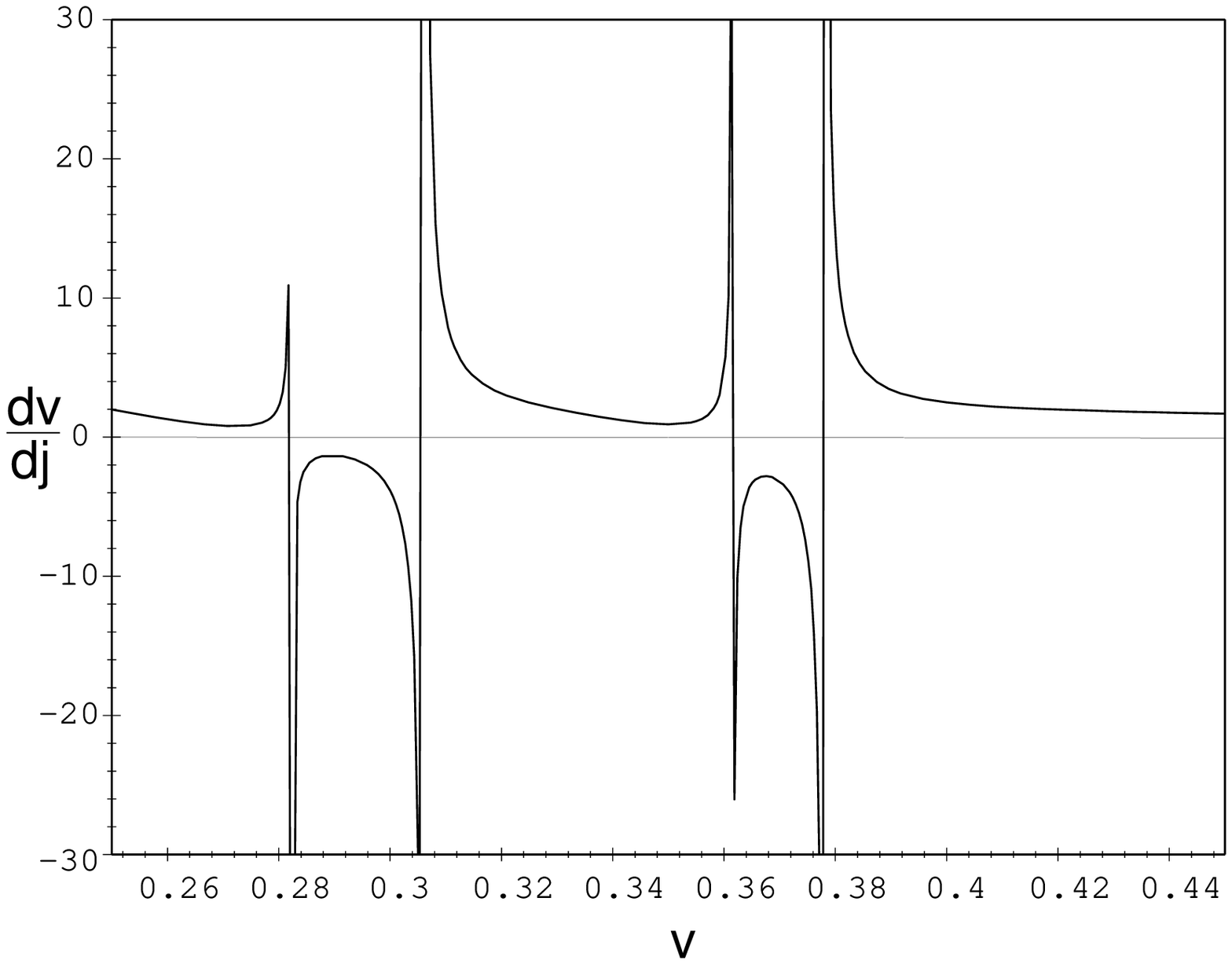}
\caption{\label{diffres} Analytical solution for the differential resistivity for 
realistic TBCCO-parameters given in fig.\ 3.} 
\end{center}
\end{figure}

In addition to this, an analytical expression for the differential resistivity 
${dv} / {dj}$ can be derived, which is plotted in fig. \ref{diffres}.
Note that there exists a region of negative differential resistivity 
for voltages $v$ slightly larger than $\Omega_{\rm LO}$, which cannot 
be reached in a current-biased experiment with a continously increasing
(decreasing) bias current. The divergence of the differential resistivity 
at the subgap structures might be connected with experimental indications for 
a significantly enhanced noise production near the maxima of the 
$I$-$V$-characteristic.

\vspace{-0.5cm}

\vspace{-.5cm}


The numerical solution of the differential equations (\ref{rsjext}) and
(\ref{oscillren}) with a Runge-Kutta-algorithm is in excellent agreement with 
the analytical results given here, but it is difficult to investigate the 
region of the $I$-$V$-characteristic with negative differential resistivity
with this method. 

From general results for small  capacity $\beta_c$ \cite{barone} 
 it is to be expected that the maximal 
current $j_{\rm max}$ and the voltage $v_{\rm max}$  of the peak 
is reduced in the presence of thermal noise. Consequently, the  
vanishing of the derivative $j^{\prime}(v)$ near $\Omega_{\rm LO}$ 
is rarely observed in experiments or numerical calculations. 

In order to reproduce the experimental data in full detail, the shape of the 
quasiparticle characteristic $I_{\rm qp}$ has to be  modified. 
Instead of the ohmic term $I_{\rm qp} = \sigma_0 {\dot \gamma}$ 
in equ. (\ref{rsjmicro}) an exponential behaviour 
\begin{equation}
I_{\rm qp} ( {\dot \gamma} ) = \exp \left( \frac{   {\dot \gamma} -1  }{v_b}
\right)
\end{equation}
 for BSCCO and a semiconductor-like dependence
\begin{equation}
I_{\rm qp} ( {\dot \gamma} ) = \frac{2 {\dot \gamma} }{1 + 
        \exp ( \frac{1- {\dot \gamma}}{v _b} ) }
\end{equation}
for TBCCO have been used with an appropriate value of $v_b$ successfully. 
The main consequence is a non-linear background 
$I_{\rm qp} ( v)$ in the $I$-$V$-characteristic, 
while the modifications of the 
quasiparticle conductivity $\sigma (v)$  are negligible. The shape and the 
position of the subgap structures are almost independent of the 
choice of $I_{\rm qp} ( {\dot \gamma} )$ and the parameter $v_b$ can be 
determined very well from parts of the $I$-$V$-characteristic away from
the subgap structures.

For an appropriate choice of parameters these analytical and numerical 
results are in excellent agreement with the experimental data, as can 
be seen in figures ({\ref{TBCCOkennlinie}) and (\ref{BSCCOkennlinie}).
All  experimental investigations of phonons in 
TBCCO \cite{phononTBCCO} and BSCCO \cite{phononBSCCO} agree that 
 infrared active c-axis phonons are observed in the
frequency range of the subgap structures. 
Theoretical calculations \cite{kulkarni} show that the
 dispersion in c-direction is small, which corresponds to a small 
ratio $\rho / (\Omega_{\rm LO} - \Omega_{\rm TO})$ in our model. 
But considerable  discrepancies in the published phonon data do not allow us 
to use the values given 
in the literature directly as input parameters. This is also due to the fact 
that usually only transversal frequencies are determined
in optical experiments or model calculations,
 while the subgap 
structures should rather  be compared with longitudinal branches. 
Also the precise  values of the McCumber parameter $\beta_c$ and the coupling 
constant $\lambda$ in TBCCO and BSCCO are unknown.
Therefore the experimental $I$-$V$-characteristics have been  fitted with the 
unconstrained parameters $\beta_c, \Omega_i, \lambda_i, \rho_i$, which turns 
out to be quite a sensitive method for the determination of these quantities, 
as an optimal fit seems to be possible only in a very restricted parameter 
range. 
As shown above the position of the peaks is given by the zeros of the phonon
dielectric constant, which is only dependent on the TO-frequencies 
$\Omega_i$ and the ratios $\lambda_i / \beta$. The absolute values 
of $\lambda_i$ and $\beta_c$ can be used to tune the relative 
strength of the peaks, 
while the values of $\rho_i$ are important for the overall curvature of the 
structure. 

The best fits for TBCCO and BSCCO are given in figures 
(\ref{TBCCOkennlinie}) and (\ref{BSCCOkennlinie}). 
The prediction for the eigenfrequencies $\nu_{\rm LO} =
\omega_{\rm LO} ({\vec k}= 0 ) / (2 \pi) $
of the longitudinal optical phonons with lowest frequencies
in both materials are in ordinary units: 
$\nu_{\rm LO,1} { =} 3.65  \, {\rm THz}$ and
$\nu_{\rm LO,2} { =} 4.70  \,{\rm THz}$ for TBCCO with a characteristic 
voltage $V_c = 27.1 {\rm mV}$
and 
$\nu_{\rm LO,1} { =} 2.96 \,{\rm THz}$, 
 $\nu_{\rm LO,2} { =} 3.90 \,{\rm THz}$ and
$\nu_{\rm LO,3} { =} 5.71 \,{\rm THz}$ for BSCCO with $V_c = 21.8 {\rm mV}$. 
These values are  compatible with the results of most optical 
experiments \cite{phononTBCCO,phononBSCCO}. 
\begin{figure}
\leavevmode
\vspace{-1.1cm}
\begin{center}
\epsfxsize=0.5\textwidth
\epsfbox{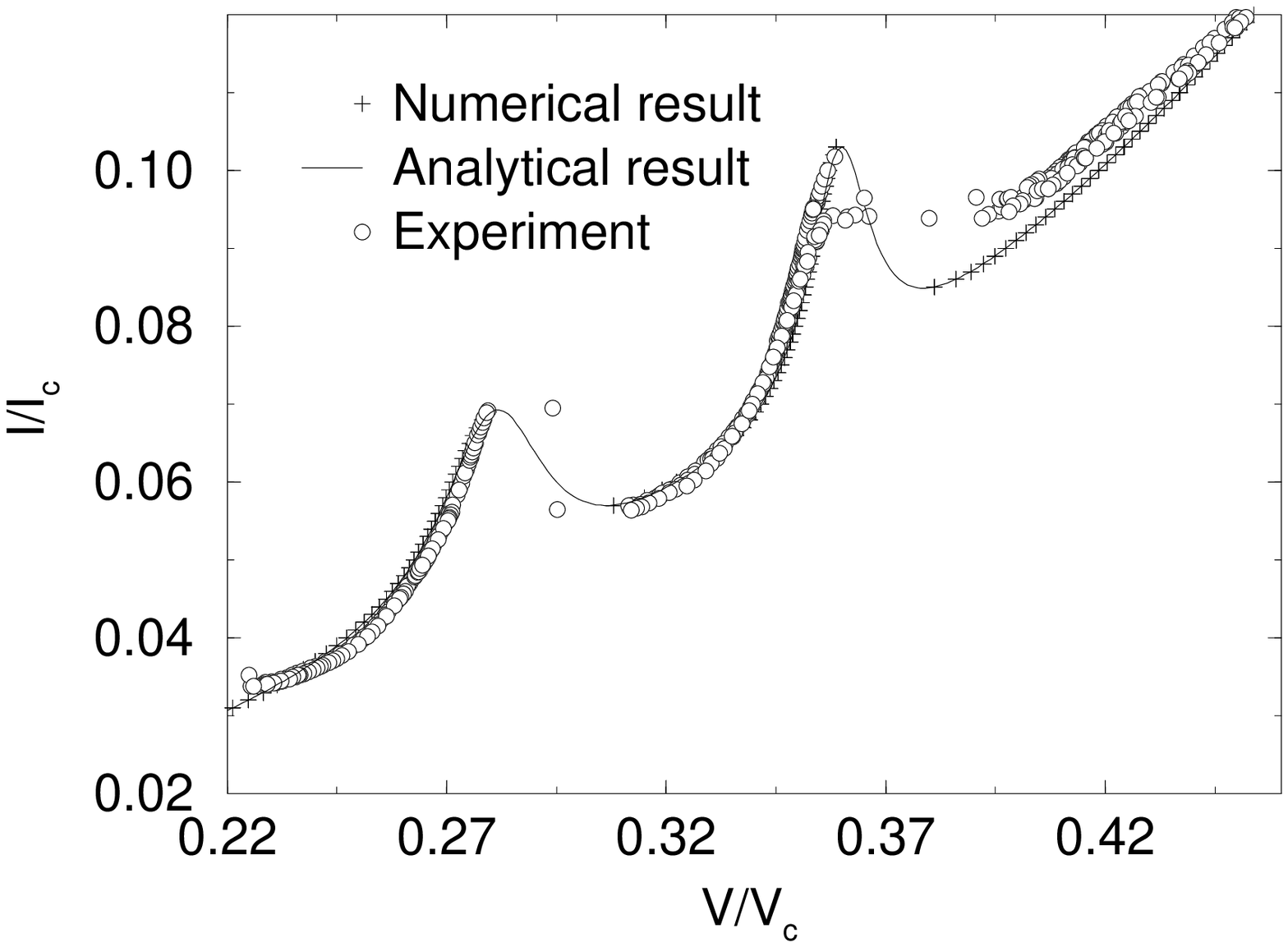}
\caption{\label{TBCCOkennlinie}\mbox{Experimental, analytical and numerical}  
        $I$-$V$-cha\-rac\-ter\-istic of TBCCO for $\beta_c=375$, $v_b=0.29$,
        $\Omega_1 = 0.25$, 
        $\lambda_1=8$, $\rho_1=0.03$, $\Omega_2=0.34$, $\lambda_2=3.5$, 
        $\rho_2=0.015$. }  
\end{center}
\end{figure}
\begin{figure}
\leavevmode
\vspace{-2.3cm}
\begin{center}
\epsfxsize=0.5\textwidth
\epsfbox{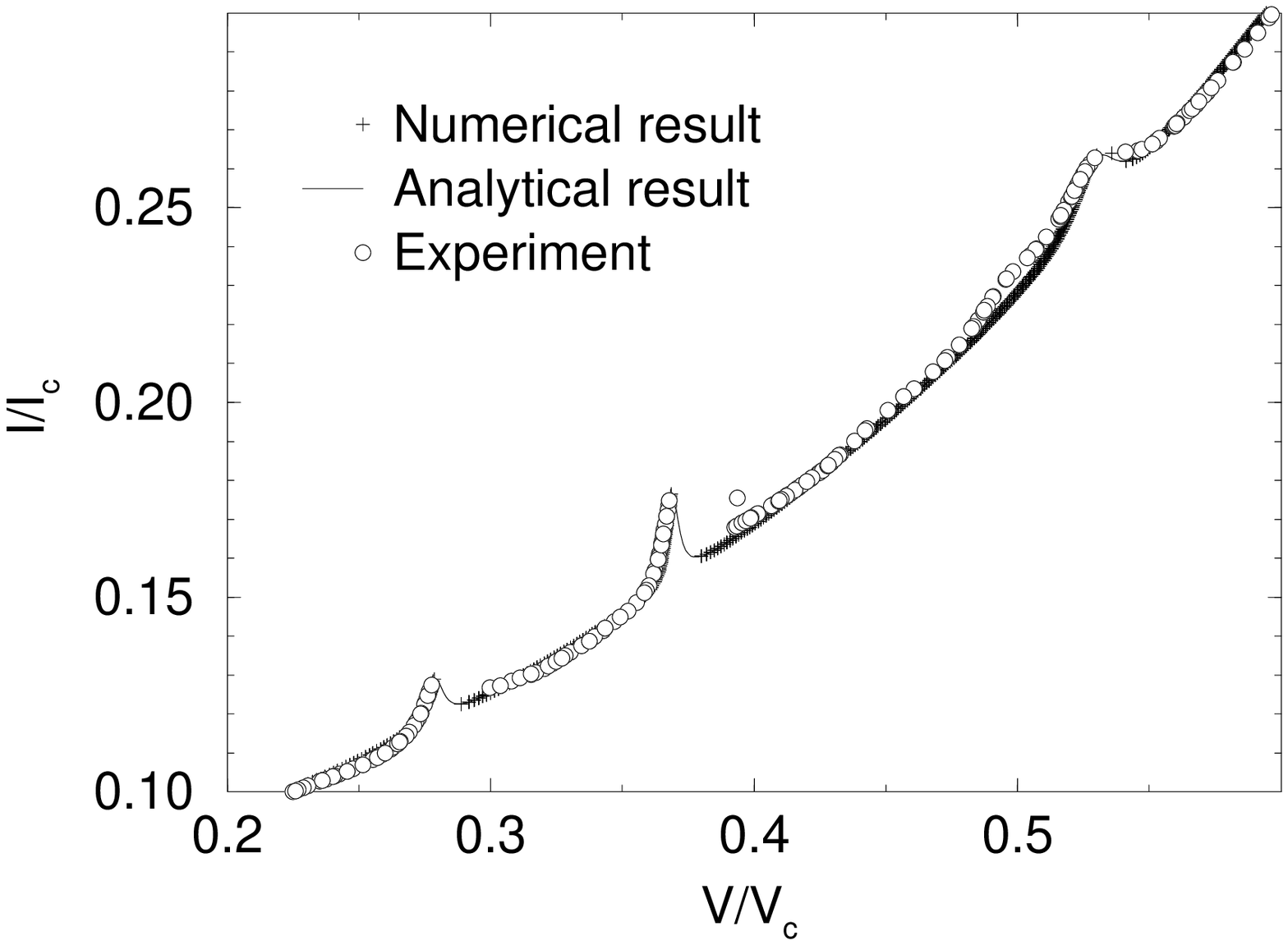}
\caption{\label{BSCCOkennlinie}\mbox{Experimental, analytical and numerical}  
         $I$-$V$-cha\-rac\-ter\-istic of BSCCO for $\beta_c=800$, $v_b=0.337$, 
        $\Omega_1 = 0.267$, 
        $\lambda_1=10$, $\rho_1=0.008$, $\Omega_2=0.345$, $\lambda_2=19$, 
        $\rho_2=0.0045$,
        $\Omega_3=0.468$, $\lambda_3=39$, $\rho_3=0.02$. } 
\end{center}
\vspace{-0.4cm}
\end{figure}

To summarize,
the recently discovered subgap structures in the $I$-$V$-characteristic 
in the intrinsic Josephson-effect in high-$T_c$-superconductors are explained 
by a coupling of Josephson oscillations 
to optical c-axis phonons within a modified RSJ-model. 
This is -- to the know\-ledge of the authors -- 
the first time that a detailed theoretical explanation of this 
effect has been given. 
Apart from perfect reproduction of the experimental data for both 
BSCCO and TBCCO  new  insights in the physical interpretation
of the peak structures are obtained.
Above all, the peak position is identified with 
the eigenfrequency of the longitudinal optical phonon, which  provides
a natural explanation for the crucial, experimental result of the
complete independence of the peak position on temperature, magnetic 
field and the geometry of the probe. 
In contrast to this, the position of the subgap structures 
is expected to depend on pressure and the isotopes contained in the material.
This fact also suggests a new approach 
for a direct measurement of this quantity, which is 
usually hard to determine in optical experiments.
It also turned out that the width of the structure is closely connected with 
the LO-TO-splitting of the phonon branch.
Also the observed proportionality of the intensity of the 
resonance to the critical current can be understood within the model presented 
above.

In principle, similar structures in the current-voltage characteristic 
can be expected near the zeros of the corresponding dielectric constant, 
if other kinds of excitations with a dipole moment are coupled to Josephson 
oscillations between the layers in a similar way.  
Extensions of this work to a microscopic theory of the Josephson effect
within the tunneling Hamiltonian formalism  in the 
presence of phonon bands $\Omega ( k_z)$ 
are currently being investigated. This  might help to 
clarify the reason for the stability of a {\em  local} oscillation 
despite of the finite coupling of oscillating ions in neighbouring junctions.

{\it Acknowledgment:} This work has been supported by a grant 
of the Bayerische Forschungsstiftung  within the research program 
FORSUPRA and  by the Studienstiftung des Deutschen Volkes (C.H.).

\end{document}